\documentclass[11pt]{article}
\pdfoutput=1

\usepackage{jheppub}
\usepackage{bigints}
\usepackage{rotating}

\usepackage{tabularx}

\usepackage{tikz}
\usepackage{fancybox}
 
\usepackage{tikz-cd}
\usepackage{ae}
\usepackage{url}

\usepackage{subfig}

\usepackage{tabularx}

\usepackage{graphicx}
\usepackage{dsfont}
\usepackage{setspace}
\usepackage{amsfonts,amsmath,amsthm,amssymb,amsbsy}
\usepackage{longtable}
\usepackage{array}
\usepackage{color}
\usepackage{needspace}
\usepackage[numbers]{natbib}
\usepackage{colortbl}

\usepackage{mmacells}

\definecolor{color_variable}{rgb}{0.1,0.55,0.25}
\definecolor{color_function}{rgb}{0.1,0.35,0.75}

\newcommand{\mmaFuncName}[1]{\text{\tt{\color{color_function}#1}}}
\makeatletter
\newcommand{\mmaFunc}[2]{\def\funcdata{}\foreach \data in {#2} {\protected@xappto\funcdata{{\tt[\,\data\,]}\!}}\mmaFuncName{#1}\funcdata\,}
\makeatother
\newcommand{\mmaFuncDef}[2]{\mmaFunc{#1}{#2}{\tt:}}
\newcommand{\mmaVar}[1]{\text{\tt{\color{color_variable}{\sl#1}}}}
\newcommand{\mmaVarDef}[1]{\mmaVar{#1\rule[-1.05pt]{7.5pt}{.75pt}}}

\usepackage{enumitem}

\usepackage{fancyhdr}

\usepackage{tabularx}

\usepackage{graphicx}
\usepackage{dsfont}
\usepackage{setspace}
\usepackage{amsfonts,amsmath,amsthm,amssymb,amsbsy}
\usepackage{longtable}
\usepackage[latin1]{inputenc}
\usepackage{array}
\usepackage{color}
\usepackage{needspace}
\usepackage[numbers]{natbib}
\usepackage{colortbl}

\usepackage{shuffle}

\usepackage{xcolor}
\definecolor{colourO}{HTML}{4FEEBB}
\definecolor{colourL}{HTML}{5DB197}
\definecolor{colourG}{HTML}{3E7CEF}
\definecolor{colourK}{HTML}{3535E3}

\usepackage{mathtools}

\usepackage{enumitem}

\usepackage{fancyhdr}

\usepackage{booktabs}
\usepackage{tcolorbox}
\usepackage{tabularx}

\usepackage{tensor}
\usepackage{mathtools}
\usepackage{blkarray, bigstrut}
\usepackage{soul}

\title{On the geometry of the orthogonal momentum amplituhedron}

\author{Tomasz \L ukowski,}\emailAdd{t.lukowski@herts.ac.uk}
\author{Robert Moerman,}\emailAdd{r.moerman@herts.ac.uk}
\author{and Jonah Stalknecht}\emailAdd{j.stalknecht@herts.ac.uk}

\affiliation{Department of Physics, Astronomy and Mathematics, \\ University of Hertfordshire, \\  Hatfield, Hertfordshire, AL10 9AB, United Kingdom}

\abstract{In this paper we study the orthogonal momentum amplituhedron $\mathcal{O}_k$, a recently introduced positive geometry that encodes the tree-level scattering amplitudes in ABJM theory. We generate the full boundary stratification of $\mathcal{O}_k$ and show that its boundaries can be labelled by so-called orthogonal Grassmannian forests (OG forests). We also determine the generating function for enumerating boundaries according to their dimension and show that the Euler characteristic of $\mathcal{O}_k$ equals one. This provides a strong indication that the orthogonal momentum amplituhedron is homeomorphic to a ball. This paper is supplemented with the Mathematica package \texttt{orthitroids} which contains useful functions for studying the positive orthogonal Grassmannian and the orthogonal momentum amplituhedron.}

\begin{document}
	
\maketitle


\section{Introduction}
In recent years we have seen tremendous interest in positive geometries \cite{Arkani-Hamed:2017tmz} that encode observables in quantum field theories, and in particular the ones that can be employed to study scattering amplitudes. A particularly fruitful theory where positive geometries can be defined is $\mathcal{N}=4$ super Yang-Mills, where the amplituhedron $\mathcal{A}_{n,k}$ \cite{Arkani-Hamed:2013jha} and the momentum amplituhedron $\mathcal{M}_{n,k}$ \cite{Damgaard:2019ztj} have been introduced to encode the tree-level scattering amplitudes. Both of these geometries are defined as the image of the positive Grassmannian through a linear map. More recently, a similar construction has been proposed for ABJM theory tree-level scattering amplitudes \cite{Huang:2021jlh,He:2021llb} using the orthogonal Grassmannian and its positive part. The resulting geometry, denoted as $\mathcal{O}_{k}$ in this paper, is $(2k-3)$-dimensional and can be thought of as a deformation of the ABHY associahedron $A_{2k-3}$ \cite{Arkani-Hamed:2017mur} where faces corresponding to even-particle planar Mandelstam variables are squashed to lower dimensional boundaries. 

In this paper we study properties of $\mathcal{O}_{k}$, and in particular provide a complete classification of its boundaries. To this end we use the algorithm developed in \cite{Lukowski:2019kqi} where it was successfully applied to find all boundaries of the amplituhedron $\mathcal{A}_{n,k}^{(2)}$, and which has been subsequently used to find all boundaries of the momentum amplituhedron $\mathcal{M}_{n,k}$ in \cite{Lukowski:2020bya}. Applying this algorithm to the orthogonal momentum amplituhedron $\mathcal{O}_k$, we observe that all boundaries can be labelled by a particular class of graphs, which we call  orthogonal Grassmannian forests, that correspond to all possible factorizations of ABJM amplitudes. This observation is analogous to the one that has been made for $\mathcal{N}=4$ sYM in \cite{Lukowski:2020bya}, where the boundaries of the momentum amplihedron $\mathcal{M}_{n,k}$ can be labelled using Grassmannian forests, see \cite{Moerman:2021cjg}. Both form a subset of the Grassmannian graphs introduced in \cite{Postnikov:2018jfq}. In this paper we summarise our explorations of the boundaries for $\mathcal{O}_k$ for $k\leq 7$, and provide a conjecture on the boundary stratification for all $k$. In particular, using the methods developed in \cite{Moerman:2021cjg}, we provide a generating function for the number of boundaries of a given dimension and this allows us to show that the Euler characteristic for the orthogonal momentum amplituhedron equals one. This story parallels the one developed for the momentum amplituhedron. 

Moreover, it has been shown in \cite{He:2021llb} that both the interior of the ABHY associahedron $A_{2k-3}$ and the interior of the orthogonal momentum amplituhedron $\mathcal{O}_k$ are diffeomorphic to the positive part of the moduli space of $n$ points on the Riemann sphere and therefore are diffeomorphic to each other. This is however not true for their closures. Nevertheless, in this paper we show that there is a simple diagrammatic way to understand how the boundaries of the associahedron naturally reduce to the boundaries of the orthogonal momentum amplituhedron.

This paper is organised as follows. In Section \ref{sec:opos} we recall the definition of the positive orthogonal Grassmannian and collect its basic properties. In Section \ref{sec:omom-boundary-strat} we define the orthogonal momentum amplituhedron following \cite{He:2021llb,Huang:2021jlh} and find its boundary stratification. Then, in Section \ref{sec:genfun} we discuss the generating function for the number of boundaries of $\mathcal{O}_k$. Section \ref{sec:mod} provides details on the diagrammatic map from boundaries of the associahedron $A_{2k-3}$ 
to those of the orthogonal momentum amplituhedron $\mathcal{O}_k$. We provide details on the Mathematica package \texttt{orthitroids} in Appendix \ref{sec:app}.


\section{Positive Orthogonal Grassmannian}
\label{sec:opos}

Let us start by recalling a few basic facts about Grassmannian spaces and their positive parts.  The \emph{real Grassmannian} $G(k,n)$ is the space of $k$-dimensional linear subspaces of $\mathbb{R}^n$. Elements of $G(k,n)$ can be represented by $k\times n$ matrices of maximal rank, where matrices related by $\mathrm{GL}(k)$ transformations are identified. The \emph{positive Grassmannian} $G_+(k,n)$, extensively studied by Postnikov in \cite{Postnikov:2006kva}, is defined as the subset of matrices $C\in G(k,n)$ for which all ordered maximal minors are non-negative. The boundary stratification of the positive Grassmannian $G_+(k,n)$ is well known \cite{Postnikov:2006kva} and it can be described in terms of \emph{positroid  cells}\footnote{The boundary stratification of $G_+(k,n)$ can be generated and studied using the Mathematica package \texttt{positroids} \cite{Bourjaily:2012gy}.}. Positroid cells of $G_+(k,n)$ are in a bijection with various combinatorial objects, including decorated permutations of type $(k,n)$, i.e.\ permutations on $[n]=\{1,\ldots,n\}$ with $k$ anti-exceedances where fixed points are coloured either black or white, equivalence classes of reduced plabic (\emph{pla}nar \emph{bic}oloured) graphs, and, more generally, equivalence classes of reduced Grassmannian graphs \cite{Postnikov:2018jfq}.

\begin{figure}
	\centering
	\begin{minipage}{0.5\textwidth}
		\centering
		\includegraphics[scale=0.7]{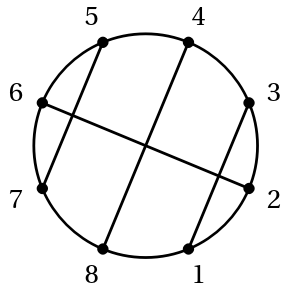}
	\end{minipage}%
	\begin{minipage}{0.5\textwidth}
		\centering
		\includegraphics[scale=0.7]{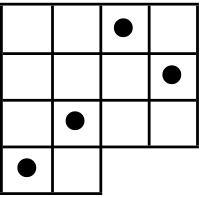}
	\end{minipage}
	\caption{A crossing diagram (left) and folded OG tableau (right) for the permutation $(1\,3)(2\,6)(4\,8)(5\,7)$.}
	\label{fig:orthitroid-example}
\end{figure}

In this paper we will be interested in an equivalent description of (a particular analytic continuation of) the \emph{positive orthogonal Grassmannian} $OG_+(k)$ introduced in \cite{Huang:2013owa}. This is defined as a $\frac{k(k-1)}{2}$-dimensional slice of the positive Grassmannian $G_+(k,2k)$ satisfying certain orthogonality conditions:
\begin{equation}
OG_+(k)=\left\{C\in G_+(k,2k):C \cdot \eta \cdot C^T=0\right\}\,,
\end{equation} 
where $\eta\coloneqq \mathrm{diag}(-,+,-,\cdots,+)$.  Similar to the positive Grassmannian, the boundary stratification of the positive orthogonal Grassmannian $OG_+(k)$ was given in \cite{Kim:2014hva} in terms of cells which we call \emph{orthitroid cells}. Orthitroid cells of $OG_+(k)$ are similarly in a bijection with various combinatorial objects, including permutations on $[2k]$ given as the product of $k$ disjoint two-cycles, a special class of marked Young diagrams called \emph{folded OG tableaux} \cite{Kim:2014hva}, and equivalence classes of reduced orthogonal Grassmannian graphs. A permutation on $[2k]$ given as the product of $k$ disjoint two-cycles can be naturally depicted as a \emph{crossing diagram} as generated by the function \texttt{oposPermToCrossing}; see Figure \ref{fig:orthitroid-example} (left) for an example. For a precise definition of folded OG tableaux, we refer the reader to Section 4.1 of \cite{Kim:2014hva}. The folded OG tableau of an orthitroid cell can be generated using the function \texttt{oposPermToYoungReducedNice}; see Figure \ref{fig:orthitroid-example} (right) for an example. Our primary interest, however, is in orthogonal Grassmannian graphs.  

We define an \emph{orthogonal Grassmannian graph} or \emph{OG graph} of \emph{type} $\underline{k}$ to be a finite planar graph $\Gamma$ with vertices $\mathcal{V}(\Gamma)$ and edges $\mathcal{E}(\Gamma)$, embedded in a disk, with $2k$ boundary vertices $\mathcal{V}_\text{ext}(\Gamma)=\{b_1,\ldots,b_{2k}\}$ of degree $1$ on the boundary of the disk labelled counterclockwise. Moreover, all internal vertices $\mathcal{V}_\text{int}(\Gamma)=\mathcal{V}(\Gamma)\setminus\mathcal{V}_\text{ext}(\Gamma)$ must be connected to the boundary of the disk via a path in $\Gamma$ and have an even degree larger than $2$. We denote by $\mathcal{E}_\text{int}(\Gamma)$ the set of \emph{internal edges} in $\Gamma$, i.e.\ those which are not adjacent to boundary vertices, and by $\Gamma_\text{int}$ the \emph{internal subgraph} of $\Gamma$. In this paper we are primarily interested in OG graphs which are forests, i.e.\ have no internal cycles. We define them in analogy with Grassmannian forests defined in \cite{Moerman:2021cjg}. An \emph{orthogonal Grassmannian forest} or \emph{OG forest} is an acyclic OG graph and an \emph{orthogonal Grassmannian tree} or \emph{OG tree} is a connected acyclic OG graph. Given an OG forest $\Gamma$ we denote the set of OG trees in $\Gamma$ by $\text{Trees}(\Gamma)$.

\begin{figure}
	\centering
	\includegraphics[scale=0.7]{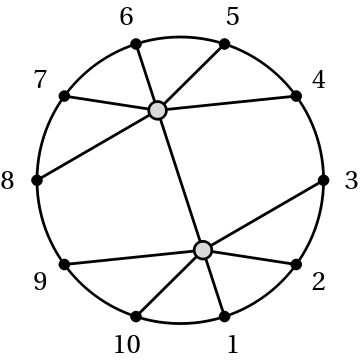}
	\caption{Example of an OG forest of type $\underline{5}$ with strand permutation $(1\,6)(2\,9)(3\,10)(4\,7)(5\,8)$.}
	\label{fig:forest-example}
\end{figure}

To each OG graph $\Gamma$, we can naturally assign a permutation in the following way: one can define a \emph{one-way strand} $\sigma_\Gamma$ which is a directed walk along edges of $\Gamma$ which starts and ends at some boundary vertices. It is defined according to the \emph{rules-of-the-road} as given in \cite{Postnikov:2018jfq}. For each vertex $v\in\mathcal{V}_\text{int}(\Gamma)$ with adjacent edges labelled clockwise $e_1,\ldots,e_d$, where $d=\deg(v)$, if $\sigma_\Gamma$ enters $v$ through edge $e_i$, it leaves through edge $e_j$, where $j=i+d/2$ (mod $d$). Said differently, if $v$ is thought of as a roundabout with $d=\deg(v)$ exits, $\sigma_\Gamma$ always exits the roundabout along the edge directly opposite the one it entered on\footnote{Consequently, each OG graph is a Grassmannian graph where each internal vertex $v$ has \emph{helicity} $h(v)=\deg(v)/2$ (c.f.\ Definitions 4.1 and 4.5 of \cite{Postnikov:2018jfq}).}. An OG graph $\Gamma$ is said to be \emph{reduced} if
\begin{enumerate}
	\item There are no strands forming closed loops in $\Gamma_\text{int}$.
	\item All strands in $\Gamma$ are simple curves without self-intersections.
	\item Any pair of strands $\alpha\ne\beta$ cannot have a \emph{bad double crossing} where there is a pair of vertices $u\ne v$ such that both $\alpha$ and $\beta$ pass from $u$ to $v$; one does allow double crossings where $\alpha$ passes from $u$ to $v$ and $\beta$ passes from $v$ to $u$.
\end{enumerate}
Importantly, OG forests are automatically reduced. The \emph{strand permutation} of a reduced OG graph of type $\underline{k}$ is a permutation $\sigma_\Gamma:[2k]\to[2k]$ where $\sigma_\Gamma(i)=j$ if the one-way strand which starts at the boundary vertex $b_i$ ends at the boundary vertex $b_j$. The strand permutation is always a product of $k$ disjoint two-cycles and it is an invariant of a reduced OG graph.  There is also a natural equivalence relation for reduced OG graphs: two reduced OG graphs $\Gamma\ne \Gamma'$ are said to be \emph{equivalent} if their strand permutations are the same, i.e.\ $\sigma_\Gamma=\sigma_{\Gamma'}$. OG forests, however, are always unique with respect to this equivalence relation. The OG forest corresponding to a permutation on $[2k]$ given as the product of $k$ disjoint two-cycles can be generated using the function \texttt{omomPermToForest}. Figure \ref{fig:forest-example} shows an example of an OG forest with its strand permutation.

As studied in \cite{Kim:2014hva}, orthitroid cells of $OG_+(k)$ are in bijection with permutations $\sigma$ on $[2k]$ given as the product of $k$ disjoint two-cycles, and therefore with equivalence classes of reduced OG graphs $\Gamma$ of type $\underline{k}$. Consequently, these combinatorial objects provide unambiguous labels for orthitroid cells in $OG_+(k)$. We denote the orthitroid cell labelled by $\sigma$ (resp., $\Gamma$) as $S_\sigma$ (resp., $S_\Gamma$). The \emph{top cell} (i.e.\ top-dimensional orthitroid cell) of $OG_+(k)$ is labelled by the permutation $(1,k+1)(2,k+2)\ldots(k,2k)$.

Each orthitroid cell in $OG_+(k)$ can be parametrized using the amalgamation method used in \cite{Huang:2013owa}, and further explored in \cite{Kim:2014hva}. In what follows we outline an algorithm, implemented in the function \texttt{oposPermToMat}, for constructing an orthogonal and positive matrix $C(\sigma)$ parametrising the orthitroid cell $S_\sigma$. This algorithm closely follows the \emph{BCFW construction} given in \cite{Kim:2014hva} with some minor modifications\footnote{We were unable to obtain a positive parametrization of orthitroid cells using the methods described in \cite{Kim:2014hva}. In particular, their BCFW construction produces complex matrices for some orthitroid cells.}. 

Let $\sigma=(i_1,j_1)(i_2,j_2)\ldots(i_k,j_k)$ with $i_\ell<j_\ell$ and $i_1<i_2<\ldots<i_k$ be a permutation labelling an orthitroid cell $S_\sigma$. The labels $i_1,\ldots,i_k$ (resp., $j_1,\ldots,j_k$) are called the \emph{pivots} (resp., \emph{sinks}) of $\sigma$ where $i_\ell$ is said to be the pivot corresponding to the sink $j_\ell$ in $\sigma$ for each $\ell=1,\ldots,k$. There is a unique zero-dimensional orthitroid cell $S_{\sigma_{0}}$ in the boundary stratification of $S_\sigma$ labelled by a permutation $\sigma_0$ which has the same pivots as $\sigma$. We call $\sigma_0$ the \emph{zero-permutation} corresponding to $\sigma$ and this can be found using the function \texttt{oposZeroPerm}. The matrix $C(\sigma_0)$ for $S_{\sigma_0}$ is constructed according to the prescription for \emph{bottom cells} given in \cite{Kim:2014hva}: $C(\sigma_0)$ is defined to be a sparse $k\times 2k$ matrix whose only non-zero entries are $C_{\ell i_\ell}(\sigma_0)=1$ and $C_{\ell j_\ell}(\sigma_0)=(-1)^{(j_\ell-i_\ell-1)/2}$ for pivots $i_\ell$ and sinks $j_\ell$ of $\sigma_0$. By construction, $C(\sigma_0)$ is real, orthogonal and positive. 

\begin{figure}
	\centering
	\begin{align*}
	\begin{gathered}
	\includegraphics[scale=0.7]{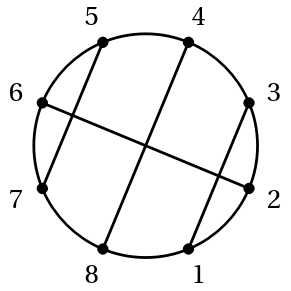}
	\end{gathered}
	\xrightarrow{\tau_3=(67)}
	\begin{gathered}
	\includegraphics[scale=0.7]{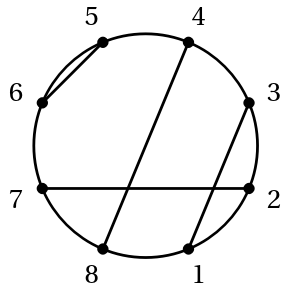}
	\end{gathered}
	\xrightarrow{\tau_2=(78)}
	\begin{gathered}
	\includegraphics[scale=0.7]{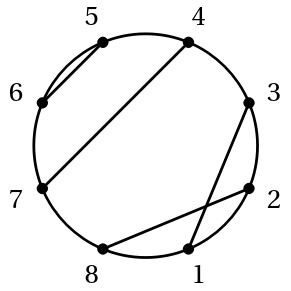}
	\end{gathered}
	\xrightarrow{\tau_1=(38)}
	\begin{gathered}
	\includegraphics[scale=0.7]{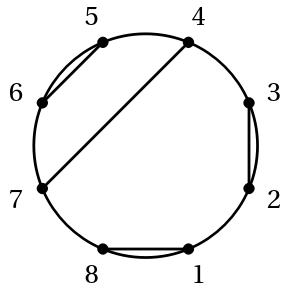}
	\end{gathered}
	\end{align*}
	\caption{BCFW construction for writing $\sigma=(1\,3)(2\,6)(4\,8)(5\,7)$ as $\sigma=\tau_3\circ\tau_2\circ\tau_1\circ\sigma_0$ where $\sigma_0=(1\,8)(2\,3)(4\,7)(5\,6)$.}
	\label{fig:bcfw}
\end{figure}

More generally, to construct the matrix $C(\sigma)$, we first find a series of transpositions $\tau_1,\ldots,\tau_d$ where $d=\dim(S_\sigma)$ such that 
\begin{align}\label{eq:transpositions}
\sigma=\tau_d\circ\tau_{d-1}\circ\cdots\circ\tau_1\circ\sigma_0\,.
\end{align}
These transpositions are obtained using the following recursive algorithm. Assume that we have already found decompositions \eqref{eq:transpositions} for all codimension-one boundaries $S_{\sigma'}\in\partial S_\sigma$. For each such $\sigma'$ we can write $\sigma=\tau_{\sigma'}\circ\sigma'$ where $\tau_{\sigma'}=(j_{\sigma',1},j_{\sigma',2})$ is a transposition that corresponds to a \emph{BCFW bridge}. To guarantee that the constructed matrix is real we need $j_{\sigma',1}+j_{\sigma',2}$ to be odd, and furthermore we want to use a BCFW bridge that acts on sinks and not on sources. The latter requirement can be translated into the condition that the line connecting the labels $j_{\sigma',1}$ and $j_{\sigma',2}$ does not cross any other line in the OG graph associated with $S_\sigma$. We have checked that it is always possible to find such a BCFW bridge for all orthitroid cells in $OG_k$ for $k\leq 7$. The algorithm described above is demonstrated in Figure \ref{fig:bcfw} for the permutation $(1\,3)(2\,6)(4\,8)(5\,7)$.

Having determined the transpositions in \eqref{eq:transpositions}, we then construct a \emph{BCFW rotation matrix} $R(\tau_m)\in SO(2k,2k)$ for each transposition $\tau_m$ according to the prescription given in \cite{Kim:2014hva}. Here
\begin{align}
SO(2k,2k)\coloneqq\left\{R\in\mathbb{R}^{2k\times 2k}:R\cdot\eta\cdot R^T=\eta\right\}\,.
\end{align}
For each transposition $\tau_m=(j_{\ell_1},j_{\ell_2})$, where $j_{\ell_1}<j_{\ell_2}$, define $R(\tau_m)$ to have the non-trivial $2\times2$ submatrix
\begin{align}
&R_{j_{\ell_1}j_{\ell_1}}(\tau_m)=R_{j_{\ell_2}j_{\ell_2}}(\tau_m)=\cosh\theta_m\,,\\ &R_{j_{\ell_1}j_{\ell_2}}(\tau_m)=R_{j_{\ell_2}j_{\ell_1}}(\tau_m)=(-1)^{(j_{\ell_2}-j_{\ell_1}-1)/2}\sinh\theta_m\,,
\end{align}
and to be the identity matrix everywhere else. It is easy to check that $R(\tau_m)$ is orthogonal. Our earlier requirement that $j_{\ell_1}+j_{\ell_2}$ be odd is now justified: it ensures that $R(\tau_m)$ is real! Finally, the matrix $C(\sigma)$ for $S_{\sigma}$ is constructed as
\begin{align}
C(\sigma)=C(\sigma_0)R(\tau_1)R(\tau_2)\cdots R(\tau_d)\,,
\end{align}
and is automatically real and orthogonal. 

This algorithm for $C(\sigma)$ is implemented in the function \texttt{oposPermToMat} and we have checked that the matrices are positive when $\theta_i$ is positive for up to $k = 5$. We believe that this will remain true for higher $k$ as well.

\section{Orthogonal Momentum Amplituhedron and its Boundary Stratification}\label{sec:omom-boundary-strat}


The orthogonal momentum amplituhedron $\mathcal{O}_k$ was recently introduced by the authors of \cite{Huang:2021jlh} and \cite{He:2021llb}\footnote{What we refer to as the `orthogonal momentum amplituhedron' was named the `ABJM momentum amplituhedron' in \cite{He:2021llb}, and they denoted it by $\mathcal{M}^{\text{3d}}(k,2k)$.} independently. The canonical form of $\mathcal{O}_k$ conjecturally encodes tree-level scattering amplitudes of ABJM theory with reduced supersymmetry in a similar way to how the canonical form of the momentum amplituhedron encodes tree-level scattering amplitudes in $\mathcal{N}=4$ super Yang-Mills theory \cite{Damgaard:2019ztj}. 
The orthogonal momentum amplituhedron $\mathcal{O}_k$ is defined as the image of the map
\begin{flalign}\label{eq:Ok-def}
	\widetilde\Phi_\Lambda \colon OG_+(k) &\to G(k,k+2)\\\nonumber
	c_{\alpha i} &\mapsto Y_\alpha^A \coloneqq c_{\alpha i} \Lambda_i^A\,, 
\end{flalign}
where $\alpha= 1,\ldots,k, \,i=1,\ldots,2k$ and $A=1,\ldots,k+2 $. Moreover, $\Lambda$ is a positive $(2k)\times (k+2)$ matrix whose entries are constrained to lie on the moment curve: $\Lambda_i^A=x_i^{A-1}$, for generic $x_1<x_2<\cdots<x_{2k}$. Importantly, it was shown in \cite{Huang:2021jlh,He:2021llb} that the image of the map $\tilde\Phi_\Lambda$ is not full-dimensional, and instead it lives inside a codimension-three subspace defined by the momentum conservation constraint
\begin{flalign}
	\sum_{i=1}^{2k} (-1)^i \big( Y^\perp \Lambda^T\big)_i^\alpha \big( Y^\perp \Lambda^T\big)_i^\beta=0\,.
\end{flalign}
The orthogonal momentum amplituhedron $\mathcal{O}_k$ is therefore $(2k-3)$-dimensional. It was also conjectured that the combinatorics of $\mathcal{O}_k$ are independent of the particular choice of positive matrix $\Lambda$, as long as it lives on the moment curve. Consequently, we will omit explicit references to $\Lambda$ in what follows. 

In this paper we determine the complete stratification of boundaries for the orthogonal momentum amplituhedron $\mathcal{O}_k$ using the same recipe employed in \cite{Ferro:2020lgp} for the momentum amplituhedron $\mathcal{M}_{n,k}$. Two main ingredients for this recipe are (1) computing the orthogonal momentum amplituhedron dimension for each orthitroid cell and (2) determining all codimension-one boundaries of the orthogonal momentum amplituhedron. These ingredients can then be combined using the algorithm introduced in \cite{Lukowski:2019kqi}. 

Let us start by defining the orthogonal momentum amplituhedron dimension for each cell in the positive orthogonal Grassmannian. Given an orthitroid cell $S_\sigma$ (resp. $S_\Gamma$) of $OG_+(k)$, we denote its image through the $\widetilde\Phi$ map as $\widetilde{\Phi}_\sigma^\circ=\widetilde{\Phi}(S_\sigma)$ (resp. $\widetilde{\Phi}_\Gamma^\circ=\widetilde{\Phi}(S_\Gamma)$) and refer to $\widetilde{\Phi}_\sigma^\circ$ (resp., $\widetilde{\Phi}_\Gamma^\circ$) as a \emph{stratum} of $\mathcal{O}_k$. We denote its closure by $\widetilde{\Phi}_\sigma$ (resp., $\widetilde{\Phi}_\Gamma$) and its dimension by $\dim(\widetilde{\Phi}_\sigma^\circ)$ (resp., $\dim(\widetilde{\Phi}_\Gamma^\circ)$). The stratum associated with the top-cell of $OG_+(k)$ is denoted by $\mathcal{O}_k^\circ$ and its dimension is $\dim(\mathcal{O}_k^\circ)=\dim(\mathcal{O}_k)=2k-3$. 

Secondly, following \cite{Huang:2021jlh}, we define the planar Mandeslstam variables
\begin{flalign}
	S_{i,i+1,\ldots,i+p}\coloneqq \sum_{i\leq j_i<j_2\leq i+p} (-1)^{j_1+j_2+1} \langle Yj_1j_2\rangle^2,
\end{flalign}
where $\langle Y i j\rangle\coloneqq \epsilon_{l_1,l_2,\ldots,l_k,l_{k+1},l_{k+2}} Y_1^{l_1} Y_2^{l_2}\cdots Y_k^{l_k} \Lambda_i^{l_{k+1}}\Lambda_j^{l_{k+2}}$. It has been conjectured that the planar Mandelstam variables are positive for all $Y\in \mathcal{O}_k$. Moreover, it is easy to check that the codimension-one boundaries of $\mathcal{O}_k$ correspond to planar Mandelstam invariants involving an odd-number of particles. On the other hand, the limit when even-particle planar Mandelstam invariants vanish corresponds to higher codimension boundaries.  

Knowing the dimension of each stratum of $\mathcal{O}_k$ as implemented in the function \texttt{omomDimension}, and all codimension-one boundaries of $\mathcal{O}_k$, we were able to find the boundary stratification of $\mathcal{O}_k$ for $k\le7$ using the algorithm from \cite{Lukowski:2019kqi}. A succinct summary of the algorithm from \cite{Lukowski:2019kqi} for the momentum amplituhedron is given in \cite{Moerman:2021cjg}. Very simply, given an orthitroid cell $S_\sigma$ of $OG_+(k)$, we say that $\widetilde{\Phi}_\sigma^\circ$ is a \emph{boundary stratum} or \emph{boundary} of $\mathcal{O}_k$ if $\widetilde{\Phi}_\sigma^\circ\cap\mathcal{O}_k^\circ=\emptyset$ and for every orthitroid cell $S_{\sigma'}\ne S_{\sigma}$ whose closure contains $S_\sigma$, $\dim(\widetilde{\Phi}_{\sigma'}^\circ)>\dim(\widetilde{\Phi}_\sigma^\circ)$. When applied to the orthogonal momentum amplituhedron, we find that the boundary stratification of $\mathcal{O}_k$ is a subposet of the orthitroid stratification of $OG_+(k)$.

\begin{figure}[h!]
\begin{center}
\raisebox{0.9cm}{\includegraphics[scale=0.4]{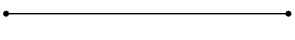}}\qquad\qquad
\includegraphics[scale=0.4]{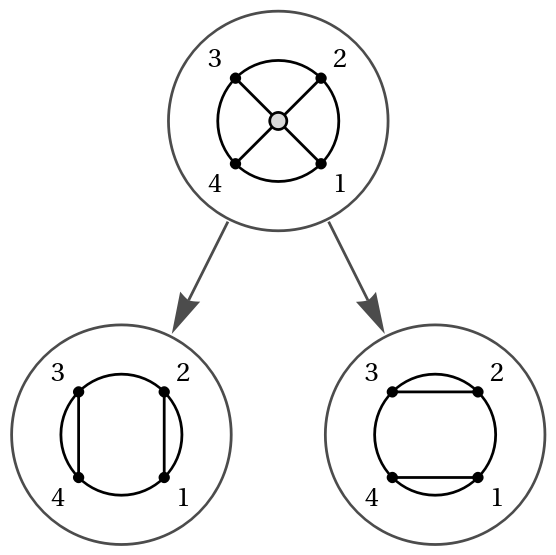}
\end{center}
\caption{The orthogonal momentum amplituhedron $\mathcal{O}_2$ and its boundary poset.}
\label{fig:k2}
\end{figure}

\begin{figure}
\begin{center}
\includegraphics[scale=0.3]{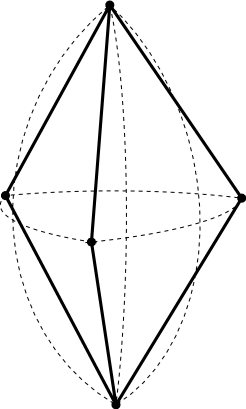}\qquad\qquad\includegraphics[scale=0.3]{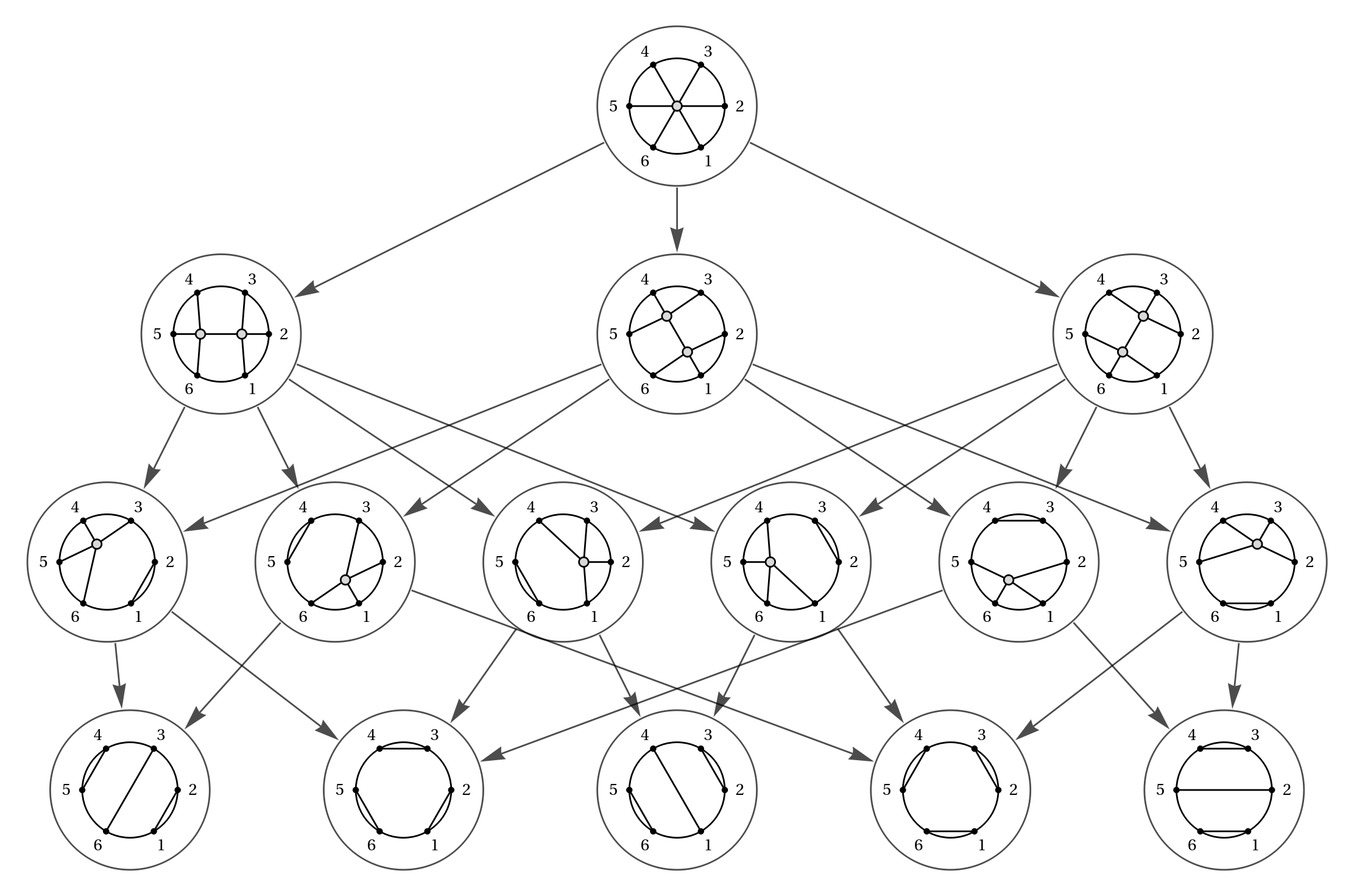}
\end{center}
\caption{The orthogonal momentum amplituhedron $\mathcal{O}_3$ and its boundary poset.}
\label{fig:k3}
\end{figure}

It was observed in \cite{Ferro:2020lgp}, and later clarified in \cite{Moerman:2021cjg}, that the boundaries of the momentum amplituhedron $\mathcal{M}_{n,k}$ are in bijection with (contracted) Grassmannian forests of type $(k,n)$. Remarkably, we found that an analogous statement is true for the orthogonal momentum amplituhedron: boundaries of the orthogonal momentum amplituhedron $\mathcal{O}_k$ are in bijection with OG forests of type $\underline{k}$. More precisely, $\widetilde{\Phi}_{\Gamma}^\circ$ is a boundary of $\mathcal{O}_k$ if and only if $\Gamma$ is an OG forest of type $\underline{k}$. We have explicitly verified this characterization for the boundaries of $\mathcal{O}_k$ in all the cases that we have studied. Starting from the simplest example, one finds that the orthogonal momentum amplituhedron $\mathcal{O}_2$ is a segment, and its boundaries are labelled as in Figure \ref{fig:k2}.
A less trivial example is the three-dimensional $\mathcal{O}_3$ (which is equal to $OG_+(3)$); its shape and boundary poset are depicted in Figure \ref{fig:k3}. 
A glimpse into the structure of the $k=4$ orthogonal momentum amplituhedron is given in Table \ref{tbl:Ok-boundaries}, where we have listed all (up to cyclic relabelling) OG forests which appear in the boundary stratification of $\mathcal{O}_4$. The complete boundary poset for $\mathcal{O}_k$ for $k=4,5,6,7$ can be generated using the \texttt{orthitroids} package, see appendix \ref{sec:app}.
\begin{table}
	\centering
	\begin{tabular}{c|c|c}
		\toprule
		$\mathrm{dim}_\mathcal{O}$ &  $\mathcal{O}_4$ & number of boundaries\\
		\midrule
		\midrule
		\raisebox{2\totalheight}0  &  \includegraphics[scale=0.4]{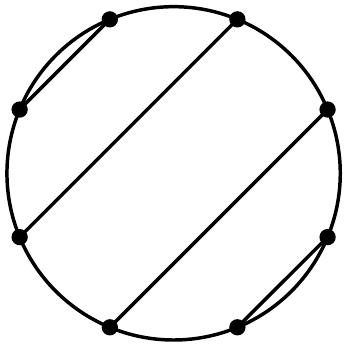} \includegraphics[scale=0.4]{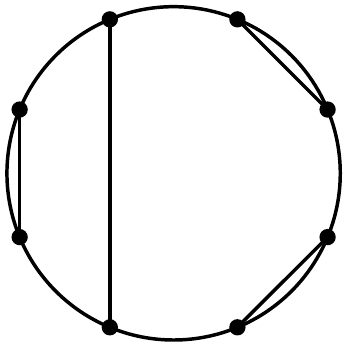} \includegraphics[scale=0.4]{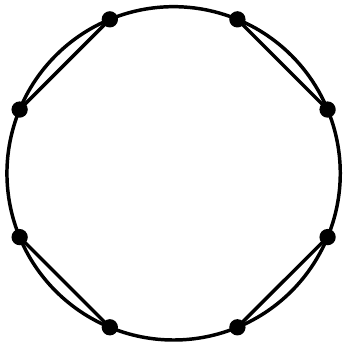} &\raisebox{2\totalheight}{4+8+2=14}\\\hline 
		\raisebox{2\totalheight}1  &  \includegraphics[scale=0.4]{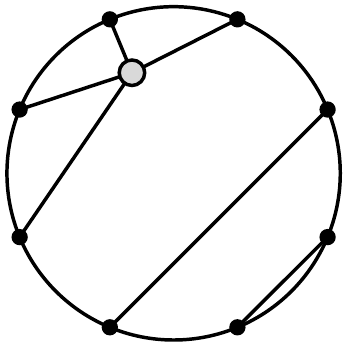} \includegraphics[scale=0.4]{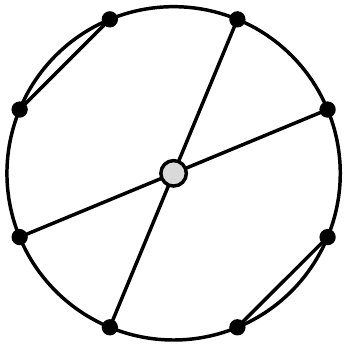} \includegraphics[scale=0.4]{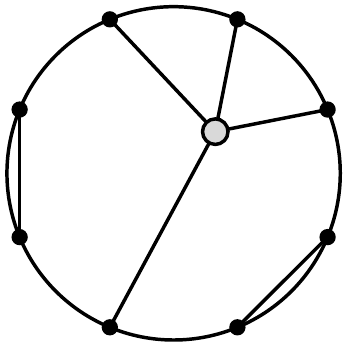} \includegraphics[scale=0.4]{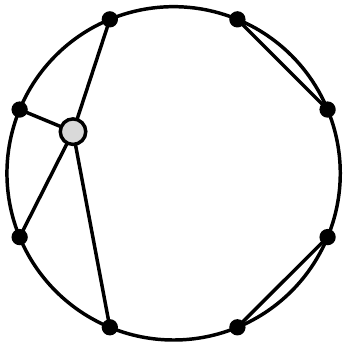}&\raisebox{2\totalheight}{8+4+8+8=28} \\\hline
		\raisebox{2\totalheight}2 &    \includegraphics[scale=0.4]{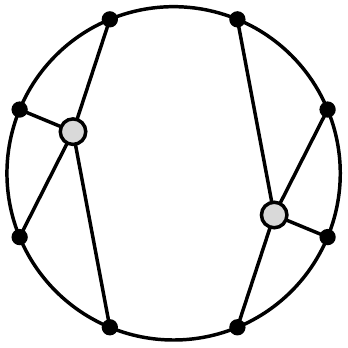} \includegraphics[scale=0.4]{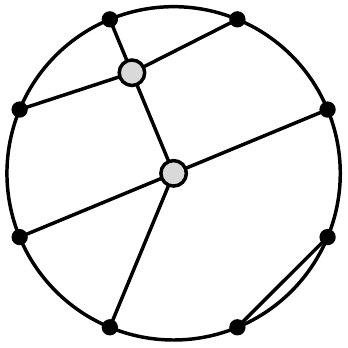} \includegraphics[scale=0.4]{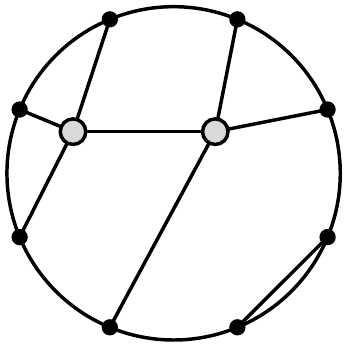} \includegraphics[scale=0.4]{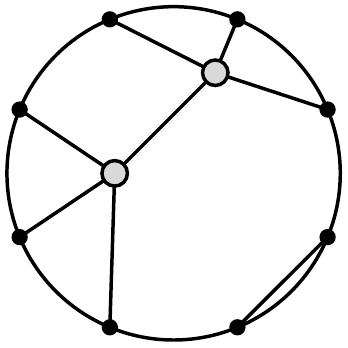} &\raisebox{2\totalheight}{4+8+8+8=28}\\\hline
		\raisebox{2\totalheight}3 &\includegraphics[scale=0.4]{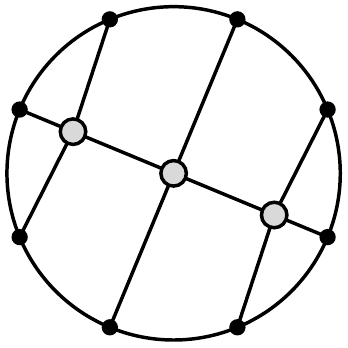} \includegraphics[scale=0.4]{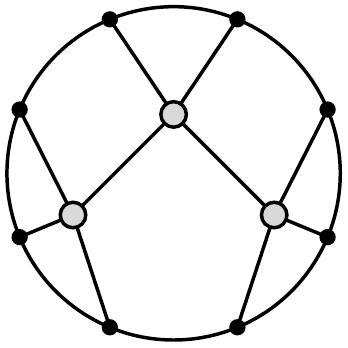} \includegraphics[scale=0.4]{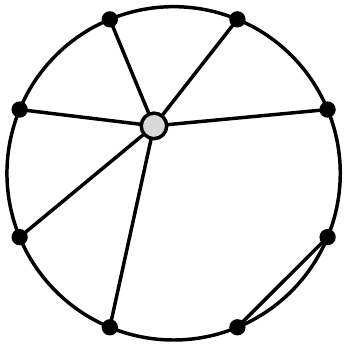}&\raisebox{2\totalheight}{4+8+8=20} \\\hline 
		\raisebox{2\totalheight}4  & \includegraphics[scale=0.4]{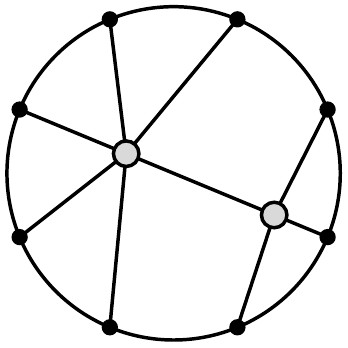}&\raisebox{2\totalheight}{8} \\\hline
		\raisebox{2\totalheight}5  & \includegraphics[scale=0.4]{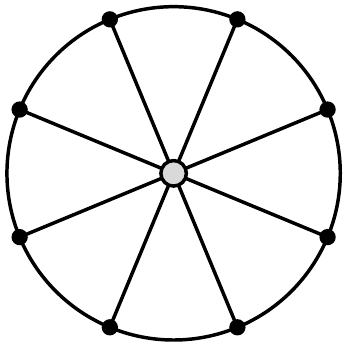} &\raisebox{2\totalheight}{1}\\
		\bottomrule
	\end{tabular}
	\caption{All types of OG forests that apear in the boundary stratification of $\mathcal{O}_4$. We find the Euler characteristic of $\mathcal{O}_4$: $\chi=14-28+28-20+8-1=1$.}
	\label{tbl:Ok-boundaries}
\end{table}

Before concluding this section, we provide an alternative way of calculating the dimension of boundaries of $\mathcal{O}_k$, i.e.\ the dimension of orthogonal momentum amplituhedron strata labelled by OG forests. Given an OG forest $\Gamma$ of type $\underline{k}$, we define $\dim_{\mathcal{O}}(\Gamma)$ by 
\begin{align}\label{eq:omom-forest-dim}
	\dim_{\mathcal{O}}(\Gamma)=\sum_{T\in\text{Trees}(\Gamma)}\dim_{\mathcal{O}}(T)\,,
\end{align}
where for each $T\in\text{Trees}(\Gamma)$
\begin{align}\label{eq:omom-tree-dim}
	\dim_{\mathcal{O}}(T)=\left\{\begin{array}{cll}		
		0&,& \text{if }|\mathcal{V}_\text{ext}(T)|=2\,,\\
		\sum\limits_{v\in\mathcal{V}_\text{int}(T)}(\deg(v)-3)&,&\text{if }|\mathcal{V}_\text{ext}(T)|>2\,.
	\end{array}\right.
\end{align}
Here $\deg(v)$ denotes the degree of the internal vertex $v$. 
We have checked that for every OG forest $\Gamma$ in $\mathcal{O}_k$ for $k\le7$ we have $\dim(\widetilde{\Phi}_\Gamma^\circ)=\dim_{\mathcal{O}}(\Gamma)$, and we believe the equality holds true for all OG forests.

Finally, we conjecture that the orthogonal momentum amplituhedron $\mathcal{O}_k$ is a CW complex with CW decomposition given by 
$\mathcal{O}_k = \sqcup_{\Gamma \in \mathcal{F}_{k}} \Phi^{\circ}_{\Gamma}$
where $\mathcal{F}_{k}$ denote the set of OG forests of type $\underline{k}$.


\section{Generating Function and Euler Characteristic}
\label{sec:genfun}

In the previous section we learned that the boundaries of the orthogonal momentum amplituhedron are labelled by OG forests. It is therefore a natural next step to enumerate all OG forests to find the $f$-vector and the Euler characteristic $\chi$ of $\mathcal{O}_k$ for all $k$. We will enumerate all OG trees and forests according to their type and orthogonal momentum amplituhedron dimension following the recipe presented in \cite{Moerman:2021cjg} where (contracted) Grassmannian trees and forests were enumerated according to their type and momentum amplituhedron dimension. In particular, for OG trees one uses the \emph{series-reduced planar tree analogue of the Exponential Formula} presented in \cite{Moerman:2021cjg}. To this end, define the statistic $f:2\mathbb{Z}_{\ge 2}\to\mathbb{Q}(q)$ which takes the degree $d\in2\mathbb{Z}_{\ge 2}$ of an internal vertex and maps it to 
$f(d) = q^{d-3}$, and let
\begin{align}
	F(x,q):=\sum_{d\in2\mathbb{Z}_{\ge 2}}f(d)x^d = q^{-3}\sum_{k=2}^{\infty}(xq)^{2k} = \frac{x^4q}{1-(xq)^2}\,,
\end{align}
be the generating function for $f$, i.e.\ the coefficient of $x^d$ in $F(x,q)$, denoted by $[x^d]F(x)$, is $f(d)$. Then using the results of \cite{Moerman:2021cjg}, the number of OG trees of type $\underline{k}$ and with orthogonal momentum amplituhedron dimension $r$ is given by $[x^{2k}q^r]\mathcal{G}_\text{tree}^{(\mathcal{O})}(x,q)$ where 
\begin{align}\label{eq:G-tree-formal}
	\mathcal{G}_\text{tree}^{(\mathcal{O})}(x,q) := x\left(x-\frac{1}{x}F(x,q)\right)^{\langle-1\rangle}_x\,,
\end{align}
and $(\ldots)^{\langle-1\rangle}_x$ denotes the compositional inverse of $(\ldots)$ with respect to the variable $x$. One can explicitly compute the compositional inverse in \eqref{eq:G-tree-formal} using the well-known \emph{Lagrange inversion formula} to find
\begin{align}
	\mathcal{G}_\text{tree}^{(\mathcal{O})}(x,q) = x^2 \left(1+\sum _{k=1}^{\infty } \sum _{\ell=1}^{\infty } \frac{1}{k} \binom{k}{\ell} \binom{2 k+\ell}{2 k+1} x^{2 k}q^{2 k-\ell}\right)\,.
\end{align} 
Then one can compute the generating function $\mathcal{G}_\text{forest}^{(\mathcal{O})}(x,q)$ which enumerates all OG forests according to their type and orthogonal momentum amplituhedron dimension using \emph{Speicher's analogue of the Exponential Formula for non-crossing partitions} \cite{Speicher}. In particular,
\begin{align}
	\mathcal{G}_\text{forest}^{(\mathcal{O})}(x,q) := \frac{1}{x}\left(\frac{x}{1+\mathcal{G}_\text{tree}^{(\mathcal{O})}(x,q)}\right)^{\langle -1\rangle}_x\,,
\end{align}
or
\begin{align}
	[x^n]\mathcal{G}_\text{forest}^{(\mathcal{O})}(x,q) := \frac{1}{n+1}[x^n]\left(1+\mathcal{G}_\text{tree}^{(\mathcal{O})}(x,q)\right)^{n+1}\,.
\end{align}
Using this formula, one can find the $f$-vector for the orthogonal momentum amplituhedron $\mathcal{O}_k$. The results for the first few values of $k$ are displayed in Table \ref{tbl:f-vector}. 
\begin{table}
	\centering
\begin{tabular}{c|l|c}
\toprule
$k$&\multicolumn{1}{c|}{$f$-vector}&$\chi$\\
\midrule
\midrule
2&$(1,2)$&1\\\hline
3&$(1,3,6,5)$&1
\\\hline
4&$(1,8,20,28,28,14)$&1
\\\hline
5&$(1,15,65,145,195,180,120,42)$&1
\\\hline
6&$(1,24,168,562,1131,1518,1430,990,495,132)$&1
\\\hline
7&$(1,35,364,1764,5019,9436,12558,12285,9009,5005,2002,429)$&1\\
\bottomrule
\end{tabular}
\caption{The $f$-vector and Euler characteristic $\chi$ of $\mathcal{O}_k$ for $k\le 7$.}
\label{tbl:f-vector}
\end{table}

One can also easily compute the Euler characteristic of $\mathcal{O}_k$ using this formula. Recall that for a CW complex, its \emph{Euler characteristic} is defined by the alternating sum $\chi = n_0-n_1+n_2 - n_3 + \ldots$ where $n_r$ denotes the number of boundaries with dimension $r$. Consequently, the Euler characteristic for $\mathcal{O}_k$ can be computed as $[x^{2k}]\mathcal{G}_\text{forest}^{(\mathcal{O})}(x,-1)$. To this end, first evaluate $\mathcal{G}_\text{tree}^{(\mathcal{O})}(x,q)$ at $q=-1$:
\begin{align}
	\mathcal{G}_\text{tree}^{(\mathcal{O})}(x,-1)=x\left(\frac{x}{1-x^2}\right)^{\langle-1\rangle}_x = -\frac{1}{2}(1+\sqrt{1+4 x^2})\,.
\end{align}
Then
\begin{align}
	\frac{x}{1+\mathcal{G}_\text{tree}^{(\mathcal{O})}(x,-1)} = \frac{2 x}{1-\sqrt{4 x^2+1}} = -\frac{1}{2x}(1+\sqrt{1+4 x^2}) = \left(\frac{x}{1-x^2}\right)^{\langle-1\rangle}_x\,,
\end{align}
and 
\begin{align}
	\mathcal{G}_\text{forest}^{(\mathcal{O})}(x,-1)=\frac{1}{x}\left(\frac{x}{1+\mathcal{G}_\text{tree}^{(\mathcal{O})}(x,q)}\right)^{\langle -1\rangle}_x = \frac{1}{x}\frac{x}{1-x^2} = \frac{1}{1-x^2}=\sum_{k=0}^{\infty}x^{2k}\,.
\end{align}
Consequently, the Euler characteristic of $\mathcal{O}_k$ is $[x^{2k}]\mathcal{G}_\text{forest}^{(\mathcal{O})}(x,-1)=1$.


\section{Diagrammatic Map between Boundaries of $\overline{\mathcal{M}}_{0,2k}^+$ and $\mathcal{O}_{k}$}\label{sec:mod}

In this section we highlight an interesting relation between the associahedron $A_{2k-3}$ and the orthogonal momentum amplituhedron $\mathcal{O}_k$. First notice that both geometries are $(2k-3)$-dimensional. Moreover, it was proposed in \cite{He:2021llb} that one can map the positive moduli space $\mathcal{M}^+_{0,2k}\coloneqq \{\sigma_1<\sigma_2<\cdots<\sigma_{2k}\}/\mathrm{SL}(2,\mathbb{R})$  directly to $\mathcal{O}_k$ using the `twistor-string map'
\begin{flalign}\label{eq:twistor-map}
	\Phi_\Lambda \colon \mathcal{M}^+_{0,2k} &\to G(k,k+2)\\\nonumber
	\sigma&\mapsto Y_\alpha^A=C_{\alpha i}(\sigma) \Lambda_i^A\,.
\end{flalign}
Here $\sigma=\{\sigma_1,\sigma_2,\cdots,\sigma_{2k}\}\in \mathcal{M}^+_{0,2k}$ is intermediately mapped to an element $C_{\alpha i}(\sigma)$ of $OG_+(k)$ through the Veronese map\begin{flalign}
	\begin{pmatrix}
		1 & 1 & \cdots & 1\\\sigma_1&\sigma_2&\cdots&\sigma_{2k}
	\end{pmatrix} \mapsto C_{\alpha i}(\sigma) =\begin{pmatrix}
		t_1 & t_2 & \cdots&t_{2k-1}&1\\
		t_1\sigma_1 & t_2\sigma_2&\cdots&t_{2k-1}\sigma_{2k-1} &\sigma_{2k}\\
		\vdots &\vdots&\ddots&\vdots&\vdots\\
		t_1\sigma_1^{k-1}&t_2\sigma_2^{k-1}&\cdots&t_{2k-1}\sigma_{2k-1}^{k-1}&\sigma_{2k}^{k-1}
	\end{pmatrix},
\end{flalign} 
with $t_i^2=(-1)^i \frac{\prod_{j\neq 2k} (\sigma_{2k}-\sigma_j)}{\prod_{j\neq i} (\sigma_{i}-\sigma_j)}$. The map \eqref{eq:twistor-map} is conjectured to provide a diffeomorphism between the interior of $\mathcal{M}^+_{0,2k}$ and the interior of $\mathcal{O}_k$. On the other hand, it is well known that the Deligne-Mumford compactification $\overline{\mathcal{M}}_{0,2k}^+$ of the positive moduli space $\mathcal{M}^+_{0,2k}$ has the boundary structure of an associahedron \cite{Deligne1969}. It is further conjectured that $\overline{\mathcal{M}}^+_{0,2k}$ is diffeomorphic to the ABHY associahedron $A_{2k-3}$ through the scattering equations \cite{Arkani-Hamed:2017mur}. Since the boundary structure of the orthogonal momentum amplituhedron differs from the one of an associahedron, it implies that the map \eqref{eq:twistor-map} does not extend to a diffeomorphism between the compactification $\overline{\mathcal{M}}_{0,2k}^+$ and the closure of $\mathcal{O}_k$. Even more well-known is the fact that the boundaries of the ABHY associahedron can be labelled by planar tree Feynman diagrams. In this section we propose a diagrammatic, partial-order-preserving map from the boundaries of the associahedron $A_{2k-3}$ to boundaries of $\mathcal{O}_{k}$ and conjecture that this extends the map \eqref{eq:twistor-map} to a homeomorphism between $\overline{\mathcal{M}}_{0,2k}^+$ and $\mathcal{O}_k$. 

\begin{figure}
	\begin{equation*}
		\vcenter{\hbox{\includegraphics[width=75mm]{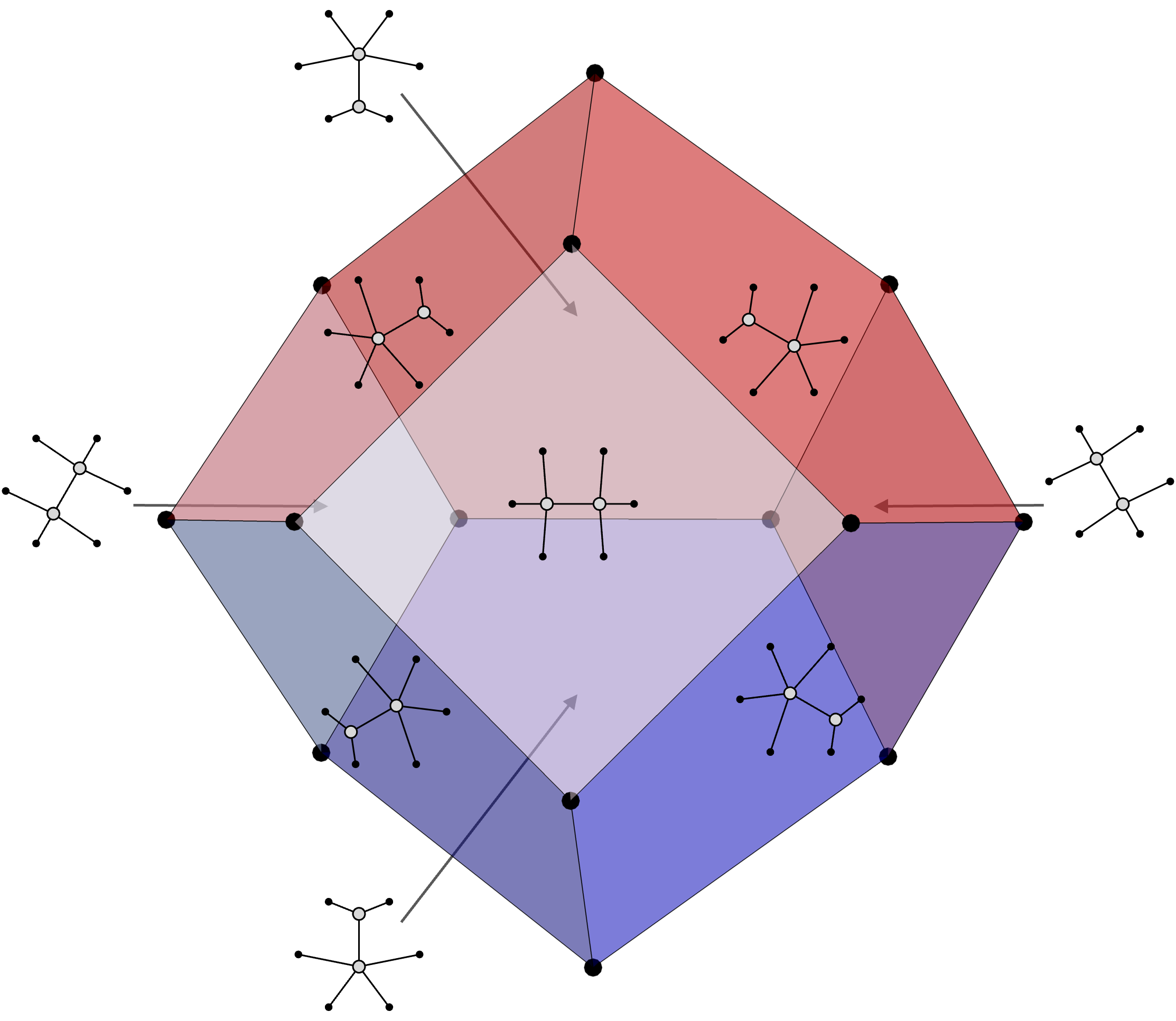}}}\quad \text{\Huge$\to$} \vcenter{\hbox{\includegraphics[width=55mm]{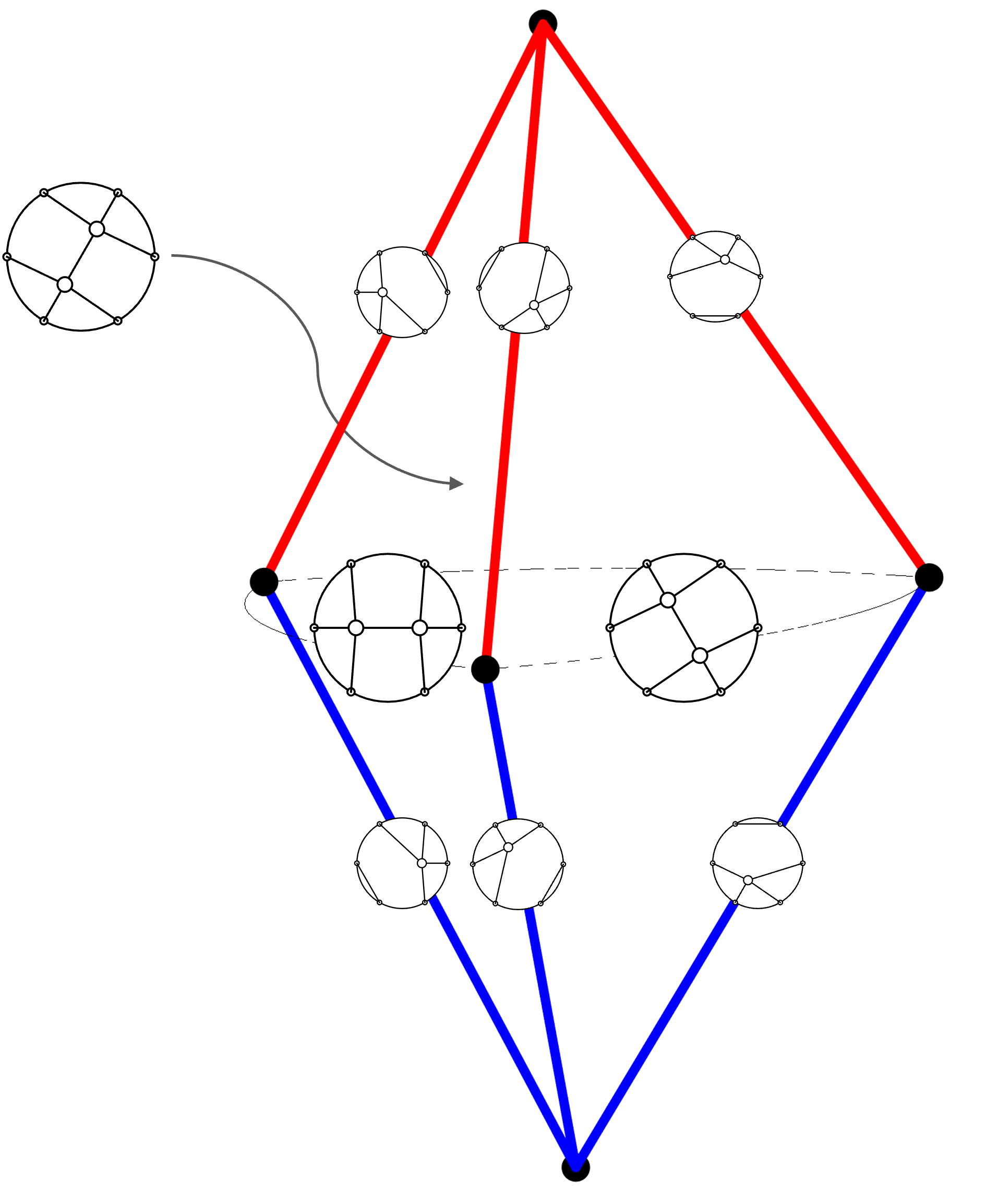}}}
	\end{equation*}
	\caption{Reduction of the three-dimensional associahedron to the orthogonal momentum amplituhedron $\mathcal{O}_3$.}
	\label{fig:ass6reduction}
\end{figure}

\begin{sidewaysfigure}
	\centering
	\includegraphics[scale=0.28]{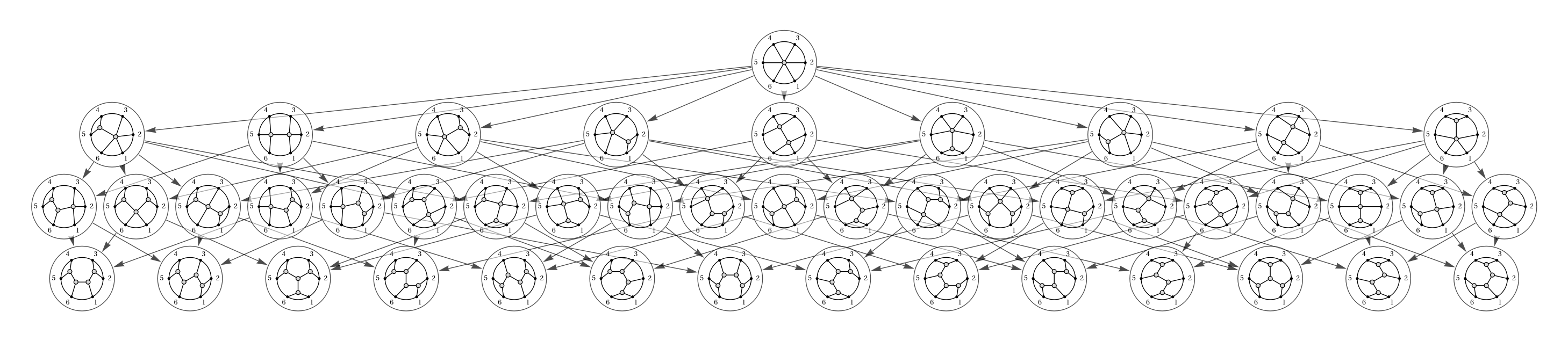}
	\includegraphics[scale=0.28]{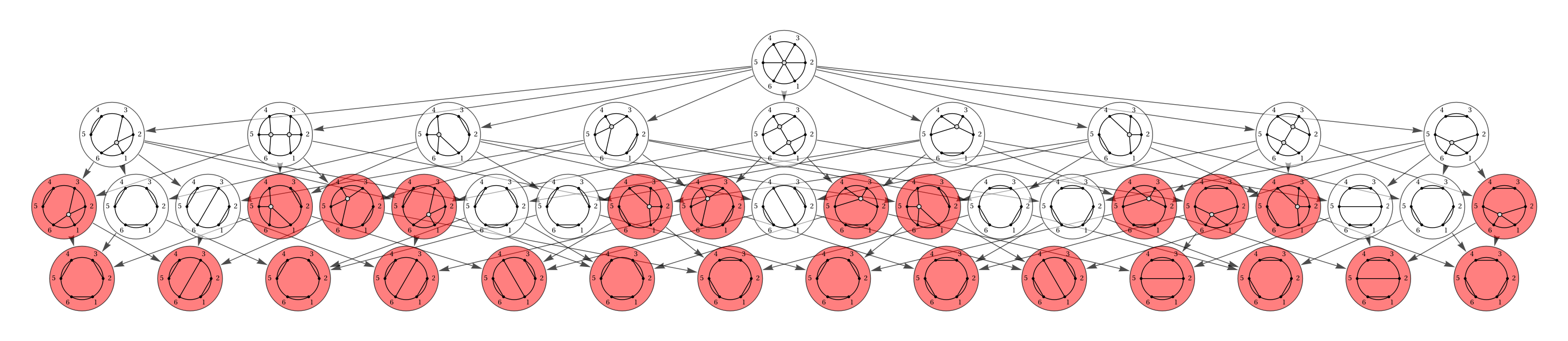}	
	\includegraphics[scale=0.28]{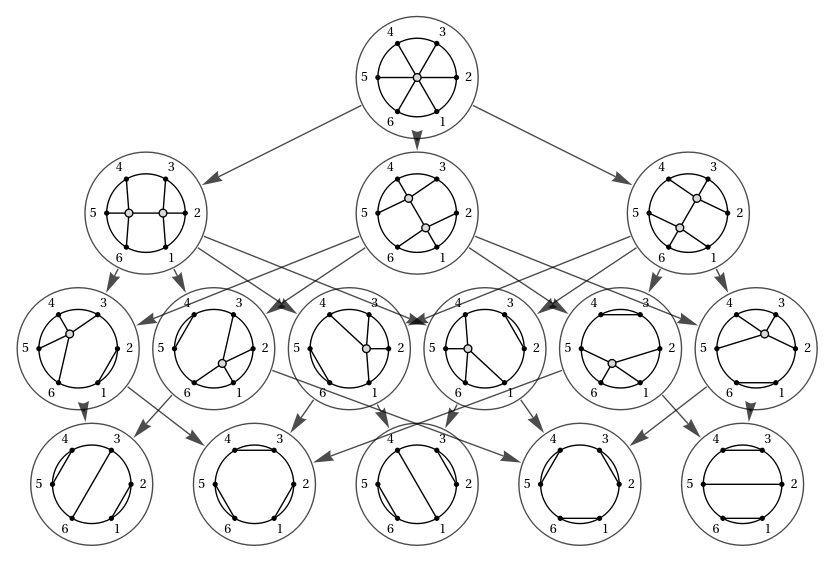}
	\caption{The top and middle Hasse diagrams depict the poset of boundaries of $A_3$ with nodes labelled by planar trees on $6$ leaves (top) and OG forests of type $3$ (middle). In the middle Hasse diagram, nodes with parents labelled by the same OG forest are coloured red. The bottom Hasse diagram is obtained from the middle one by identifying nodes labelled by  the same OG forest.}
	\label{fig:modPoset}
\end{sidewaysfigure}

Note that the Feynman diagrams appearing in the boundary stratification of $A_{2k-3}$ can be labelled by the planar Mandelstam invariants that correspond to the internal edges of the diagram\footnote{This can be represented by a set of non-intersecting chords of a $2k$-gon. The chord between vertices $i$ and $j$ corresponds to the planar Mandelstam variable $S_{i,i+1,\cdots,i+j-1}$.}. 
We now propose the following map from planar tree Feynman diagrams to OG forests: given a planar tree Feynman diagram, remove all internal edges which correspond to even-particle planar Mandelstam invariants and replace all vertices of degree $2$ with a single edge. It is not difficult to convince oneself that the resulting diagram is an OG forest. Furthermore, it is clear that every OG forest can be obtained from at least one planar tree Feynman diagram, as can be seen by starting from an OG forest and connecting all  disconnected parts by internal edges. Since all planar tree Feynman diagrams on $2k$ leaves appear in the boundary stratification of $A_{2k-3}$, and given our knowledge of the boundary stratification of $\mathcal{O}_k$ discussed in Section \ref{sec:omom-boundary-strat}, it is clear that the map defined above is surjective from boundary elements of $A_{2k-3}$ to boundary elements of $\mathcal{O}_k$. We furthermore claim that the poset induced by the boundary stratification of $A_{2k-3}$ will result in the poset corresponding to the boundary stratification of $\mathcal{O}_k$. The simplest example is the case when $k=2$, where both the associahedron $A_1$ and the orthogonal momentum amplituhedron $\mathcal{O}_2$ are segments. The first non-trivial example is the one for $k=3$ where the $f$-vector for the associahedron $A_3$ given by $(1,9,21,14)$ reduces to the $f$-vector of $\mathcal{O}_3$ given by $(1,3,6,5)$. This is depicted in Figure \ref{fig:ass6reduction} at the level of geometry and in Figure \ref{fig:modPoset} at the level of boundary posets. 
We have performed direct calculations for $k\leq 5$ and explicitly confirmed that the boundary stratification of the associahedron reduces to the boundary stratification found in Section \ref{sec:omom-boundary-strat}, and we believe this continues to be be true for all $k$.

The reduction discussed above also allows us to explain some of the statements about physical boundaries of $\mathcal{O}_k$ made in \cite{Huang:2021jlh,He:2021llb}. In particular, let us take a closer look at the codimension-one boundaries of the associahedron. Recall that the interior of the associahedron $A_{2k-3}$ can be represented by a Feynman diagram with a single degree $2k$ vertex. Going to one of its codimension-one boundaries corresponds to dissolving the degree $2k$ vertex into two vertices of degree $(2k-p+1)$ and $(p+1)$, with $2\leq p\leq 2k-2$, connected via a single edge. For odd (resp., even) $p$, the internal edge corresponds to an odd-particle (resp., even-particle) Mandelstam variable. Via the map defined above, Feynman diagrams with a single edge corresponding to an odd-particle Mandelstam variable will remain unchanged, and we thus interpret the Feynman diagram as an OG tree without changing anything. From \eqref{eq:omom-tree-dim}, the resulting OG tree will have an orthogonal momentum amplituhedron dimension of $(2k-p+1-3)+(p+1-3)=2k-4$, and therefore labels a codimension-one boundary of $\mathcal{O}_k$. On the other hand, Feynman diagrams with a single edge corresponding to an even-particle Mandelstam variable will be mapped to OG forests consisting of two disconnected OG trees, each consisting of a single vertex of degree $2k-p$ and degree $p$, respectively. From \eqref{eq:omom-forest-dim} we find that these OG trees have orthogonal momentum amplituhedron dimension $(2k-p-3)+(p-3)=2k-6$, with an exception for the cases where $p=2$ or $p=2k-2$ when the dimension of the OG forest is $2k-5$. These results agree with the discussions about the physical boundaries of $\mathcal{O}_k$ found in \cite{He:2021llb,Huang:2021jlh}.

\appendix
\section{Mathematica package \texttt{orthitroid}}
\label{sec:app}
While working on this project we have developed a Mathematica package \texttt{orthitroids} which contains many useful functions for studying the positive orthogonal Grassmannian and the orthogonal momentum amplituhedron. It parallels some of the functionality of the \texttt{positroids} package \cite{Bourjaily:2012gy} and the \texttt{amplituhedronBoundaries} package \cite{Lukowski:2020bya}. In this appendix we provide a list of some useful functions from the \texttt{orthitroids} package. We use the following three name spaces for distinguishing functions:
\begin{itemize}
\item \texttt{opos} -- for functions related to the othogonal positive Grassmannian $OG_+(k)$;
\item \texttt{omom} -- for functions related to the orthogonal momentum amplituhedron $\mathcal{O}_k$;
\item \texttt{mod} -- for functions related to the Deligne-Mumford compactification of the positive part of the moduli space of $2k$ points on the Riemann sphere $\overline{\mathcal{M}}_{0,2k}^+$.
\end{itemize}

\subsection{Positive Orthogonal Grassmannian}
\begin{itemize}
\item \mmaFuncDef{oposTopCell}{\mmaVarDef{k}} returns the permutation for the top-dimensional orthitroid cell of the positive orthogonal Grassmannian.

\item \mmaFuncDef{oposPermToCrossing}{\mmaVarDef{perm}} returns a \emph{crossing diagram} for the permutation \mmaVar{perm}.
\begin{mmaCell}[moredefined={oposPermToCrossing}]{Input}
oposPermToCrossing[\{\{1,3\},\{2,7\},\{4,6\},\{5,8\}\}]
\end{mmaCell}

\begin{mmaCell}[verbatimenv=]{Output}
\raisebox{-0.5\totalheight}{\includegraphics[scale=0.7]{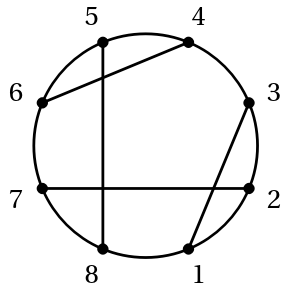}}
\end{mmaCell}

\item
\mmaFuncDef{oposPermToYoungNice}{\mmaVarDef{perm}} returns the Young diagram for the orthitroid cell labelled by the permutation \mmaVar{perm}. 

\begin{mmaCell}[moredefined={oposPermToYoungNice}]{Input}
oposPermToYoungNice[\{\{1,3\},\{2,7\},\{4,6\},\{5,8\}\}]
\end{mmaCell}

\begin{mmaCell}[verbatimenv=]{Output}
\raisebox{-0.5\totalheight}{\includegraphics[scale=0.7]{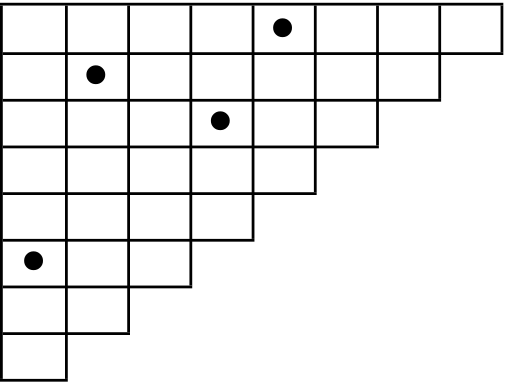}}
\end{mmaCell}

\item \mmaFuncDef{oposPermToYoungReducedNice}{\mmaVarDef{perm}} returns the reduced (folded) Young diagram for the orthitroid cell labelled by the permutation \mmaVar{perm}. 

\begin{mmaCell}[moredefined={oposPermToYoungReducedNice}]{Input}
oposPermToYoungReducedNice[\{\{1,3\},\{2,7\},\{4,6\},\{5,8\}\}]
\end{mmaCell}

\begin{mmaCell}[verbatimenv=]{Output}
\raisebox{-0.5\totalheight}{\includegraphics[scale=0.7]{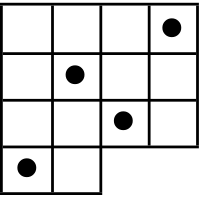}}
\end{mmaCell}

\item
\mmaFuncDef{oposDimension}{\mmaVarDef{perm}} returns the positive orthogonal Grassmannian dimension of the orthitroid cell labelled by the permutation \mmaVar{perm}.  

\begin{mmaCell}[moredefined={oposDimension}]{Input}
oposDimension[\{\{1,3\},\{2,7\},\{4,6\},\{5,8\}\}]
\end{mmaCell}

\begin{mmaCell}[]{Output}
3
\end{mmaCell}

\item
\mmaFuncDef{oposBoundary}{\mmaVarDef{perm}} returns the list of codimension-one boundaries of the orthitroid cell labelled by the permutation \mmaVar{perm}. 

\begin{mmaCell}[moredefined={oposBoundary}]{Input}
oposBoundary[\{\{1,3\},\{2,7\},\{4,6\},\{5,8\}\}]
\end{mmaCell}

\begin{mmaCell}[]{Output}
\{\{\{1,2\},\{3,7\},\{4,6\},\{5,8\}\},\{\{1,3\},\{2,5\},\{4,6\},\{7,8\}\},
\{\{1,3\},\{2,7\},\{4,5\},\{6,8\}\},\{\{1,7\},\{2,3\},\{4,6\},\{5,8\}\},
\{\{1,3\},\{2,8\},\{4,6\},\{5,7\}\},\{\{1,3\},\{2,7\},\{4,8\},\{5,6\}\}\}
\end{mmaCell}

\item
\mmaFuncDef{oposInverseBoundary}{\mmaVarDef{perm}} returns the list of orthitroid cells which have the orthitroid cell labelled by the permutation \mmaVar{perm} as a codimension-one boundary. 

\begin{mmaCell}[moredefined={oposInverseBoundary}]{Input}
oposInverseBoundary[\{\{1,3\},\{2,7\},\{4,6\},\{5,8\}\}]
\end{mmaCell}

\begin{mmaCell}[]{Output}
\{\{\{1,4\},\{2,7\},\{3,6\},\{5,8\}\},\{\{1,5\},\{2,7\},\{3,8\},\{4,6\}\},
\{\{1,3\},\{2,6\},\{4,7\},\{5,8\}\}\}
\end{mmaCell}

\item
\mmaFuncDef{oposStratification}{\mmaVarDef{perm}} returns all boundaries (of all codimensions) of the orthitroid cell labelled by the permutation \mmaVar{perm}.

\item
\mmaFuncDef{oposInverseStratification}{\mmaVarDef{perm}} returns the list of orthitroid cells which have the orthitroid cell labelled by the permutation \mmaVar{perm} in its boundary stratification. 

\item
\mmaFuncDef{oposPermToMat}{\mmaVarDef{perm}} returns a matrix parametrizing the orthitroid cell labelled by the permutation \mmaVar{perm} with $c_i=\cosh(\theta_i)$ and $s_i=\sinh(\theta_i)$.	
\begin{mmaCell}[moredefined={oposPermToMat}]{Input}
MatrixForm@oposPermToMat[\{\{1,3\},\{2,7\},\{4,6\},\{5,8\}\}]
\end{mmaCell}

\begin{mmaCell}[verbatimenv=]{Output}
$\left(
\begin{array}{cccccccc}
1 & 0 & -s_3 & 0 & 0 & 0 & -c_3 s_1 & -c_1 c_3 \\
0 & 1 & c_3 & 0 & 0 & 0 & s_1 s_3 & c_1 s_3 \\
0 & 0 & 0 & 1 & 0 & -s_2 & -c_1 c_2 & -c_2 s_1 \\
0 & 0 & 0 & 0 & 1 & c_2 & c_1 s_2 & s_1 s_2 \\
\end{array}
\right)$
\end{mmaCell}
\end{itemize}	

\subsection{Orthogonal Momentum Amplituhedron}

\begin{itemize}
\item \mmaFuncDef{omomPermToForest}{\mmaVarDef{perm}} returns the OG forest for the permutation \mmaVar{perm}.
\begin{mmaCell}[moredefined={omomPermToForest}]{Input}
omomPermToForest[\{\{1,3\},\{2,7\},\{4,6\},\{5,8\}\}]
\end{mmaCell}

\begin{mmaCell}[verbatimenv=]{Output}
\raisebox{-0.5\totalheight}{\includegraphics[scale=0.7]{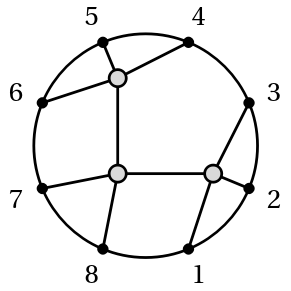}}
\end{mmaCell}
\item
\mmaFuncDef{omomDimension}{\mmaVarDef{perm}} returns the orthogonal momentum amplituhedron dimension of the cell labelled by the permutation \mmaVar{perm}. 

\begin{mmaCell}[moredefined={omomDimension}]{Input}
omomDimension[\{\{1,3\},\{2,7\},\{4,6\},\{5,8\}\}]
\end{mmaCell}

\begin{mmaCell}[]{Output}
3
\end{mmaCell}

\item
\mmaFuncDef{omomBoundary}{\mmaVarDef{perm}} returns the list of codimension-one boundaries of the orthogonal momentum amplituhedron cell labelled by the permutation \mmaVar{perm}.

\item
\mmaFuncDef{omomInverseBoundary}{\mmaVarDef{perm}} returns the list of orthogonal momentum amplituhedron cells which have the cell labelled by the permutation \mmaVar{perm} as a codimension-one boundary. 

\begin{mmaCell}[moredefined={omomInverseBoundary}]{Input}
omomInverseBoundary[\{\{1,3\},\{2,7\},\{4,6\},\{5,8\}\}]
\end{mmaCell}

\begin{mmaCell}[]{Output}
\{\{\{1,3\},\{2,6\},\{4,7\},\{5,8\}\},\{\{1,5\},\{2,7\},\{3,8\},\{4,6\}\}\}
\end{mmaCell}

\item \mmaFuncDef{omomStratification}{\mmaVarDef{perm}} returns all boundaries (of all codimensions) of the orthogonal momentum amplituhedron cell labelled by the permutation \mmaVar{perm}.

\item \mmaFuncDef{omomInverseStratification}{\mmaVarDef{perm}} returns the list of orthogonal momentum amplituhedron cells which have the cell labelled by the permutation \mmaVar{perm} in its boundary stratification. 

\end{itemize}

\subsection{Moduli Space}

\begin{itemize}
\item \mmaFuncDef{modDiagonalsToPlanarTree}{\mmaVarDef{k},\mmaVarDef{listOfDiags}} returns a planar tree on $2\mmaVar{k}$ leaves which is dual to the dissection of a regular $2\mmaVar{k}$-gon specified by the diagonals in \mmaVar{listOfDiags}.
\begin{mmaCell}[moredefined={modDiagonalsToPlanarTree}]{Input}
modDiagonalsToPlanarTree[4][\{\{1,3\},\{1,5\}\}]
\end{mmaCell}

\begin{mmaCell}[verbatimenv=]{Output}
\raisebox{-0.5\totalheight}{\includegraphics[scale=0.7]{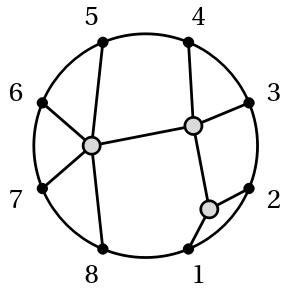}}
\end{mmaCell}

\item \mmaFuncDef{modDiagonalsToForest}{\mmaVarDef{k},\mmaVarDef{listOfDiags}} returns the corresponding OG forest of type $\underline{\mmaVar{k}}$ obtained from \mmaFunc{modDiagonalsToPlanarTree}{\mmaVar{k},\mmaVar{listOfDiags}} via the map described in this paper.
\begin{mmaCell}[moredefined={modDiagonalsToForest}]{Input}
modDiagonalsToForest[4][\{\{1,3\},\{1,5\}\}]
\end{mmaCell}

\begin{mmaCell}[verbatimenv=]{Output}
\raisebox{-0.5\totalheight}{\includegraphics[scale=0.7]{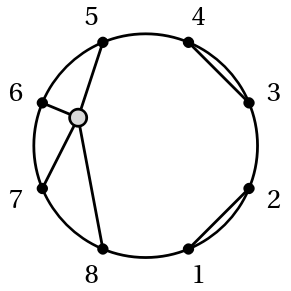}}
\end{mmaCell}
\end{itemize}

\bibliographystyle{nb}

\bibliography{lms_v1}

\end{document}